\documentclass[12pt]{article}
\usepackage{float}\restylefloat{figure}
\usepackage{epsfig}
\usepackage{graphicx}
\usepackage{epstopdf}
\def\bm#1{\mbox{\boldmath $#1$}}
\def\beq{\begin{equation}}
\def\eeq{\end{equation}}
\def\t2{\mbox{  }}
\def\rst1{\mbox{ }}

\begin{document} 
\title{\normalsize {\bf Yukawa potentials in systems with partial periodic boundary conditions II : Lekner sums for quasi-two dimensional systems.}}
\author{\normalsize Martial MAZARS\footnote{Electronic mail: Martial.Mazars@th.u-psud.fr} \\
\small Laboratoire de Physique Th\'eorique (UMR 8627),\\
\small Universit\'e de Paris Sud XI, B\^atiment 210, 91405 Orsay Cedex, FRANCE}
\maketitle
\normalsize
\begin{center}{\bf Abstract}\end{center}
Yukawa potentials may be long ranged when the Debye screening length is large. In computer simulations, such long ranged potentials have to be taken into account with convenient algorithms to avoid systematic bias in the sampling of the phase space. Recently, we have provided Ewald sums for quasi-two dimensional systems with Yukawa interaction potentials [M. Mazars, {\it J. Chem. Phys.\/}, {\bf 126}, 056101 (2007) and M. Mazars, {\it Mol. Phys.\/}, Paper I]. Sometimes, Lekner sums are used as an alternative to Ewald sums for Coulomb systems. In the present work, we derive the Lekner sums for quasi-two dimensional systems with Yukawa interaction potentials and we give some numerical tests for pratical implementations. The main result of this paper is to outline that Lekner sums cannot be considered as an alternative to Ewald sums for Yukawa potentials. As a conclusion to this work : Lekner sums should not be used for quasi-two dimensional systems with Yukawa interaction potentials.

\newpage
\section{Introduction}

This work follows a previous paper on the derivation of Ewald sums for Yukawa potentials in quasi-two-dimensional systems \cite{f1,f2}. The concern of the present work is to derive the Lekner sums for Yukawa potentials and to explore their possible applicabilities to quasi-two-dimensional systems.\\
J. Lekner has developed his method as an alternative to Ewald sums for Coulomb potential in systems with periodic boundary conditions \cite{f3,f4} ; from an analytical point of view Lekner and Ewald sums are fully equivalent. In Lekner sums, the summations over the periodic images are transformed as summations over modified Bessel functions K$_0$ (or K$_1$ for molecular dynamics implementations) ; because of the asymptotic behaviour for large values of the argument of  Bessel functions, the summations are rapidly convergent ; but, because of the singular behaviour of Bessel functions for small values of the argument, the summations converge very slowly for some particular configurations of the pair of particles \cite{f5,f6}. This last property of Lekner sums introduces some complications in numerical implementations of Lekner sums. There are mainly two methods to improve the convergence rate of these summations : the Lekner-cyclic \cite{f4,f7} and Lekner-Sperb \cite{f8,f9,f10} methods. A review of the uses of Lekner sums for Coulomb potential in computer simulations has already been done in refs.\cite{f5,f7} and would not be reproduced here.\\
In the present work, we derive Lekner sums for quasi-two dimensional systems with Yukawa interaction potentials and we give some numerical tests for pratical implementations that use the Lekner-cyclic method. The paper is organised as follow. In section 2, we derive the Lekner sums for Yukawa potentials in quasi-two dimensional systems and we show from an analytical point of view that Lekner and Ewald sums are equivalent. In section 3, we give some numerical tests for implementations of Lekner sums, these tests show that the use of Lekner sums for Yukawa potential should be avoided since to reproduce with accuracy the Coulomb limit of  Yukawa potentials, one has to modify the summations by using Taylor expansions for small screening parameter $\kappa$ that would not allow to implement Lekner sums for any value of $\kappa$. The conclusion in section 4 states that one should use Ewald sums for Yukawa potentials in quasi-two dimensional systems \cite{f1,f2}. 

\section{Lekner sums for Yukawa interactions in quasi-two dimensional systems.}

We consider a system made of $N$ particles interacting via Yukawa potentials. The simulation box have partial periodic boundary conditions, with spatial periodicity $L_x$ and $L_y$, applied respectively to directions $x$ and $y$, whereas no periodic boundary conditions are taken in the third direction $z$, parallel to the unitary vector $\hat{\bm{e}}_z$. In the simulation box, the position of the particle $i$ is defined by $\bm{r}_i=(x_i,y_i,z_i)$.\\
The particle-particle interaction energy is given by
\begin{equation}
E_{cc}(\mbox{Yukawa}; \kappa)=\frac{1}{2}\sum_{i=1}^{N}\sum_{i\neq j}^N Q_i Q_j \Phi(\bm{r}_{ij})+\frac{1}{2}\sum_{i=1}^{N}Q_i^2\Phi_0
\end{equation}
with
\begin{equation}
\displaystyle\Phi(\bm{r})=\sum_{\bm{n}}\frac{\exp(-\kappa \mid \bm{r}+\bm{n}\mid)}{\mid \bm{r}+\bm{n}\mid}\mbox{    and    } \Phi_0\displaystyle = \sum_{\bm{n}\neq 0}\frac{\exp(-\kappa \mid \bm{n}\mid)}{\mid \bm{n}\mid}
\end{equation}
where we used the condensed notations 
\begin{center}
$\displaystyle \mid \bm{r}+\bm{n}\mid = \sqrt{(x+n_xL_x)^2+(y+n_yL_y)^2+z^2}$ and $\displaystyle \mid \bm{n}\mid = \sqrt{n_x^2L_x^2+n_y^2L_y^2}$
\end{center}
with $n_x$ and $n_y$ integer numbers associated with periodic images of  the box.\\
As for the Ewald sums \cite{f1}, the lattice sum is transformed by using the integral relation
\begin{equation}
\frac{\exp(-\kappa \mid \bm{r}+\bm{n}\mid)}{\mid \bm{r}+\bm{n}\mid} = \frac{1}{\sqrt{\pi}}\int_0^{\infty}\frac{dt}{\sqrt{t}}\exp(-\frac{\kappa^2}{4t}-\mid \bm{r}+\bm{n}\mid^2 t)
\end{equation}
we obtain
\begin{equation}
\begin{array}{ll}
\displaystyle\Phi(\bm{r})&\displaystyle=\frac{1}{\sqrt{\pi}}\int_0^{\infty}\frac{dt}{\sqrt{t}}\exp(-\frac{\kappa^2}{4t}-z^2t)\mbox{\Large{[}}\sum_{n=-\infty}^{+\infty}\exp\mbox{\large{(}}-(\frac{x}{L_x}+n)^2L_x^2 t\mbox{\large{)}}\mbox{\Large{]}}\\
&\\
&\displaystyle\times\mbox{\Large{[}}\sum_{m=-\infty}^{+\infty}\exp\mbox{\large{(}}-(\frac{y}{L_y}+m)^2L_y^2 t\mbox{\large{)}}\mbox{\Large{]}}
\end{array}
\end{equation}
Then, the Poisson-Jacobi identity  
\begin{equation}
\displaystyle\sum_{n=-\infty}^{+\infty}\exp(-(n+\frac{x}{L_x})^2 L_x^2 t) = \frac{1}{L_x}\sqrt{\frac{\pi}{t}}\mbox{\Large{[}}1+2\sum_{p=1}^{\infty}\cos\mbox{\large{(}}2\pi p \frac{x}{L_x}\mbox{\large{)}}e^{-\pi^2 p^2/L_x^2 t}\mbox{\Large{]}}
\end{equation}
is applied to only one of the summations over the periodic images, it gives
\begin{equation}
\begin{array}{ll}
\displaystyle\Phi(\bm{r})&\displaystyle=\frac{1}{L_x}\sum_{m=-\infty}^{+\infty}\int_0^{\infty}\frac{dt}{t}\exp\mbox{\Large{(}}-\frac{\kappa^2}{4t}-\mbox{\large{(}}(y+mL_y)^2+z^2)t\mbox{\large{)}}\mbox{\Large{)}}\\
&\\
&\displaystyle+\frac{2}{L_x}\sum_{p=1}^{+\infty}\cos\mbox{\large{(}}2\pi p \frac{x}{L_x}\mbox{\large{)}}\sum_{m=-\infty}^{+\infty}\int_0^{\infty}\frac{dt}{t}\exp\mbox{\Large{[}}-\mbox{\large{(}}\kappa^2+\frac{4\pi^2p^2}{L_x^2}\mbox{\large{)}}\frac{1}{4t}-\mbox{\large{(}}(y+mL_y)^2+z^2\mbox{\large{)}}t\mbox{\Large{]}}
\end{array}
\end{equation}
thus, taking into account the integral
\begin{equation}
\int_{0}^{\infty}\frac{dt}{t}\exp\mbox{\Large{[}}-\frac{A^2}{4t}-B^2 t \mbox{\Large{]}}=2\mbox{ K}_0(AB)
\end{equation}
we find
\begin{equation}
\begin{array}{ll}
\displaystyle\Phi(\bm{r})&\displaystyle=\Phi(x,y,z;L_x,L_y ; \infty, \infty)=\frac{2}{L_x}\sum_{m=-\infty}^{+\infty}\mbox{ K}_0\mbox{\Large{[}}\kappa\sqrt{(y+mL_y)^2+z^2}\mbox{\Large{]}}\\
&\\
&\displaystyle+\frac{4}{L_x}\sum_{p=1}^{+\infty}\cos\mbox{\large{(}}2\pi p \frac{x}{L_x}\mbox{\large{)}}\sum_{m=-\infty}^{+\infty}\mbox{ K}_0\mbox{\Large{[}}\sqrt{\mbox{\large{(}}\kappa^2+\frac{4\pi^2p^2}{L_x^2}\mbox{\large{)}}\mbox{\large{(}}(y+mL_y)^2+z^2\mbox{\large{)}}}\mbox{\Large{]}}
\end{array}
\end{equation}
where the notation $\Phi(x,y,z;L_x,L_y ; \infty, \infty)$ indicates that the summations over $p$ and $m$ are not truncated.\\
The self contribution  $\Phi_0$ is needed to compute the particle-particle energy. This contribution reads as
\begin{equation}
\begin{array}{ll}
\displaystyle\Phi_0&\displaystyle = \Phi_0(L_x,L_y ; \infty, \infty) = \sum_{\bm{n}\neq 0}\frac{\exp(-\kappa \mid \bm{n}\mid)}{\mid \bm{n}\mid}\\
&\\
&\displaystyle = \frac{8}{L_x}\sum_{p=1}^{+\infty}\sum_{m=1}^{+\infty}\mbox{ K}_0\mbox{\huge{[}}m L_y\sqrt{\kappa^2+\frac{4\pi^2p^2}{L_x^2}}\mbox{\huge{]}}+\frac{4}{L_x}\sum_{m=1}^{+\infty}\mbox{ K}_0(m \kappa L_y)-\frac{2 }{L_x}\ln(1-\exp(-\kappa L_x))
\end{array}
\end{equation}
The particle-particle interaction energy $E_{cc}(\mbox{Yukawa}; \kappa)$ is given by  Eqs.(1), (8) and (9), for any value of the inverse screening length $\kappa$.\\
In the Coulomb limit ($\kappa\rightarrow 0$), the singular terms are the first contribution of the right handed side of Eq.(8) and the last two contributions of Eq.(9). For the singular contribution in Eq.(8), we find \cite{f10}
\begin{equation}
\begin{array}{ll}
\displaystyle\frac{2}{L_x}\sum_{m=-\infty}^{+\infty}\mbox{ K}_0\mbox{\Large{[}}\kappa\sqrt{(y+mL_y)^2+z^2}\mbox{\Large{]}}&\displaystyle=\frac{2\pi}{\kappa A}-\frac{\ln 2}{L_x}-\frac{1}{L_x}\ln\mbox{\huge{[}}\cosh\mbox{\large{(}}\frac{2\pi z}{L_y}\mbox{\large{)}}-\cos\mbox{\large{(}}\frac{2\pi y}{L_y}\mbox{\large{)}}\mbox{\huge{]}}\\
&\\
&\displaystyle +\kappa\frac{\pi z^2}{A}+o(\kappa^2)
\end{array}
\end{equation}
with $A=L_x L_y$, and for the singular terms in Eq.(9), one has
\begin{equation}
\left\{
\begin{array}{ll}
\displaystyle 4\sum_{m=1}^{+\infty}\mbox{ K}_0(m \kappa L_y) &\displaystyle = \frac{2\pi}{\kappa L_y}+2\ln \mbox{\large{(}}\frac{\kappa L_y}{4\pi}\mbox{\large{)}}+2\gamma -\frac{\kappa^2 L_y^2}{4\pi^2}\sum_{l=1}^{\infty}\frac{1}{l^3}+o(\kappa^4)\\
&\\
\displaystyle\ln(1-\exp(-\kappa L_x))&\displaystyle = \ln(\kappa L_x)-\frac{1}{2}\kappa L_x+\frac{1}{24}(\kappa L_x)^2-\frac{1}{6}(\kappa L_x)^3+o(\kappa^4)
\end{array}
\right .
\end{equation}
with $\gamma \simeq 0.5772156649...$ the Euler's constant. The first expansion in (11) have been obtained with the help of Schl\"omilch series \cite{f11}.\\ 
On Eq.(11), one should note that the singular terms as $\ln \kappa$ are cancelled by adding them in $\Phi_0$. For numerical applications, Eqs.(10) and (11) arise some difficulties in the Coulomb limit : expansions in the right hand side of these equations are exact as $\kappa\rightarrow 0$, as long as the Bessel series are not truncated. However, in applications, one has to truncate these series to an index $m_c$ and as $\kappa$ becomes smaller, $m_c$ has to be taken greater so that  to obtain a reasonable accuracy for expansion in Eqs.(10) and (11). This point will be illustrated in the next section. \\
Therefore, in the Coulomb limit, we have 
\begin{equation}
\begin{array}{ll}
\displaystyle\Phi(\bm{r})&\displaystyle=\frac{2\pi}{\kappa A}+\frac{4}{L_x}\sum_{p=1}^{+\infty}\cos\mbox{\large{(}}2\pi p \frac{x}{L_x}\mbox{\large{)}}\sum_{m=-\infty}^{+\infty}\mbox{ K}_0\mbox{\huge{[}}\frac{2\pi p}{L_x} \sqrt{(y+mL_y)^2+z^2}\mbox{\huge{]}}\\
&\\
&\displaystyle -\frac{\ln 2}{L_x}-\frac{1}{L_x}\ln\mbox{\huge{[}}\cosh\mbox{\large{(}}\frac{2\pi z}{L_y}\mbox{\large{)}}-\cos\mbox{\large{(}}\frac{2\pi y}{L_y}\mbox{\large{)}}\mbox{\huge{]}}+\kappa \frac{\pi z^2}{A}+o(\kappa^2)
\end{array}
\end{equation}
and 
\begin{equation}
\begin{array}{ll}
\displaystyle\Phi_0&\displaystyle= \frac{2\pi}{\kappa A}+\frac{8}{L_x}\sum_{p=1}^{\infty}\sum_{m=1}^{\infty}\mbox{ K}_0\mbox{\Large{[}}2\pi pm \frac{L_y}{L_x}\mbox{\Large{]}}+\frac{2}{L_x}\mbox{\Large{[}}\gamma+\ln\mbox{\Large{(}}\frac{L_y}{4\pi L_x}\mbox{\Large{)}}\mbox{\Large{]}}+\kappa+ o(\kappa^2)
\end{array}
\end{equation}
Thus, as $\kappa\rightarrow 0$, we find
\begin{equation}
\displaystyle E_{cc}(\mbox{Yukawa}; \kappa\rightarrow 0)=E_{cc}(\mbox{Coulomb})+\frac{\pi}{A}\mbox{\large{(}}\sum_i Q_i\mbox{\large{)}}^2\frac{1}{\kappa}+o(\kappa)
\end{equation}
with $E_{cc}(\mbox{Coulomb})$ the particle-particle interaction energy of a quasi-two-dimensional system of particles interacting with Coulomb potentials and computed with the Lekner sums as 
\begin{equation}
\begin{array}{ll}
\displaystyle E_{cc}(\mbox{Coulomb})&\displaystyle= \frac{2}{L_x}\sum_{i,j}\mbox{}'Q_i Q_j\sum_{p=1}^{+\infty}\sum_{m=-\infty}^{+\infty}\cos\mbox{\large{(}}2\pi p \frac{x_{ij}}{L_x}\mbox{\large{)}}\mbox{ K}_0\mbox{\huge{[}}\frac{2\pi p}{L_x} \sqrt{(y_{ij}+mL_y)^2+z_{ij}^2}\mbox{\huge{]}}\\
&\\
&\displaystyle -\frac{1}{2L_x}\sum_{i\neq j}Q_i Q_j \ln\mbox{\huge{[}}\cosh\mbox{\large{(}}\frac{2\pi z_{ij}}{L_y}\mbox{\large{)}}-\cos\mbox{\large{(}}\frac{2\pi y_{ij}}{L_y}\mbox{\large{)}}\mbox{\huge{]}}-\frac{1}{2L_x}\mbox{\large{(}}\sum_{i=1}^{N}Q_i\mbox{\large{)}}^2\ln 2\\
 &\\
&\displaystyle +\frac{1}{L_x}\mbox{\large{(}}\sum_{i=1}^{N}Q_i^2\mbox{\large{)}}\mbox{\large{[}}\gamma + \ln\mbox{\large{(}}\frac{L_y}{L_x}\mbox{\large{)}}-\frac{1}{2}\ln(8\pi^2) \mbox{\large{]}}
\end{array}
\end{equation}
where the prime in summations over the particles indicates that $m=0$ is not included if $i=j$. Equation (14) is exactly the same as the one found for Ewald sums, this outlines the consistency between the Lekner and Ewald methods for Yukawa potentials.\\
In numerical uses, summations over $p$ and $m$ in Eqs.(8) and (9) have to be truncated to some finite values $p_c$ and $m_c$, therefore, in numerical implementations of Lekner sums, for both Coulomb and Yukawa potentials, one should take care of the slow convergence of the sums over the Bessel functions for some particular configurations of the system of particles \cite{f7,f5,f6}. There are mainly two ways to counter the slow convergence behaviour of the Lekner sums, Lekner-Sperb \cite{f10} and the Lekner-cyclic methods \cite{f7}.\\
The method proposed by Sperb begins with an analytical identity similar to Eq.(8), then an alternative expression which converges faster than the sums over the Bessel function, for the configurations leading to a slow convergence behaviour is derived and used as an alternative to the slow convergence \cite{f8,f9} ; with this method, some cautions are needed to take properly into account the electroneutrality of the system \cite{f10}.\\ 
In the Lekner-cyclic method, one uses a symmetry property of the analytical forms of Lekner sums. Namely, if in Eqs.(8) and (9),  one substitutes $x$ to $y$ and $L_x$ to $L_y$ then the energy $E_{cc}(\mbox{Yukawa}; \kappa)$ has to remain unchanged (for a detailed discussion and numerical tests about Coulomb potential, see refs.\cite{f7,f5}) ; in the following, when substitutions $x\rightarrow y$ and $y\rightarrow x$ are done, the new functions obtained from Eq.(8) and (9) are noted $\Psi(x,y,z;L_x,L_y;p_c,m_c)$ and $\Psi_0(L_x,L_y;p_c,m_c)$, i.e.
\begin{equation}
\left\{
\begin{array}{l}
\Psi(x,y,z;L_x,L_y;p_c,m_c)=\Phi(y,x,z;L_y,L_x;p_c,m_c)\\[0.1in]
\Psi_0(L_x,L_y;p_c,m_c)=\Phi_0(L_y,L_x;p_c,m_c)
\end{array}
\right .
\end{equation}
and obviously, we have
\begin{equation}
\left\{
\begin{array}{ll}
\displaystyle \lim_{(p_c, m_c)\rightarrow(\infty,\infty)}\Phi(x,y,z;L_x,L_y;p_c,m_c) &\displaystyle = \lim_{(p_c, m_c)\rightarrow(\infty,\infty)}\Psi(x,y,z;L_x,L_y;p_c,m_c)\\
&\\
\displaystyle\lim_{(p_c, m_c)\rightarrow(\infty,\infty)}\Phi_0(L_x,L_y;p_c,m_c) &\displaystyle = \lim_{(p_c, m_c)\rightarrow(\infty,\infty)}\Psi_0(L_x,L_y;p_c,m_c)
\end{array}
\right .
\end{equation}
Then, to implement the Lekner-cyclic method, a criterion has to be chosen to determinate which function, $\Phi$ or $\Psi$, has to be used to compute  $E_{cc}(\mbox{Yukawa}; \kappa)$ (or $E_{cc}(\mbox{Coulomb})$). Some numerical tests to help to choose criterions for different systems will be provided in the next section. In the remaining of the present section we will focus on the singular term as $\kappa\rightarrow 0$.\\
In the Coulomb limit, the singular term in Eq.(14) is cancelled if the system of particles fulfills the electroneutrality (i.e. $\sum_{i=1}^NQ_i=0$), as in the computations done in the previous paper to obtain the Ewald sums \cite{f1}. Such electroneutrality implies that there is some attractive interactions between particles (for instance systems of macroions with opposite charge), but for most of systems in interest, as colloids or dust plasmas, interactions between the macroions are always repulsive. Thus, a neutralizing background has to be included in the systems to permit to obtain the correct Coulomb limit. In ref.\cite{f1}, several kinds of neutralizing background have been considered. In the present work we will restrict ourselves to the {\it monolayer neutralizing background\/} (system (a) of ref.\cite{f1}), since the computation for other backgrounds may be easily done following the methods below and in the previous paper.\\
The monolayer neutralizing background  is a plan with an uniform surface charge density localised at $z=0$, particles may be localised on both sides of the plan or their location may be restricted also to only one half-space. As in ref.\cite{f1}, the system considered is made by the monolayer and the particles ; for this system, the charge density in the right hand side of the Helmholtz equation is given by
\begin{equation}
\displaystyle \rho(\bm{r})=\sum_iQ_i\delta(\bm{r}-\bm{r}_i)+\sigma\delta(z)
\end{equation}
assuming that for all particles $Q_i=Q$, the electroneutrality for this system reads as
\begin{equation}
NQ+A\sigma=0
\end{equation}
Taking into account this background, the energy of the system is given by 
\begin{equation}
E(\mbox{Yukawa};\kappa)=E_{cc}(\mbox{Yukawa};\kappa)+E_{cS}(\kappa)+E_{SS}(\kappa)
\end{equation}
where $E_{cS}(\kappa)$ is the interaction energy of the particles with the neutralizing background, while $E_{SS}(\kappa)$ is the interaction energy of the neutralizing background with itself ; these interaction energies are given by
\begin{equation}
\left\{
\begin{array}{ll}
\displaystyle 
\displaystyle E_{cS}(\kappa)&\displaystyle=-\frac{NQ^2}{A}\sum_{i=1}^N\int_S d\bm{s}\int_{-\infty}^{+\infty}dz \delta(z) \sum_{\bm{n}}\frac{\exp(-\kappa \mid \bm{r}_i-\bm{r}+\bm{n}\mid)}{\mid \bm{r}_i-\bm{r}+\bm{n}\mid}\\
&\\
\displaystyle E_{SS}(\kappa)&\displaystyle=\frac{N^2Q^2}{2A^2}\int_S d\bm{s}'\int_{-\infty}^{+\infty}dz' \delta(z')\int_S d\bm{s}\int_{-\infty}^{+\infty}dz \delta(z) \sum_{\bm{n}}\frac{\exp(-\kappa \mid \bm{r}'-\bm{r}+\bm{n}\mid)}{\mid \bm{r}'-\bm{r}+\bm{n}\mid}
\end{array}
\right .
\end{equation}
then, with Eq.(3) and the Poisson-Jacobi identity Eq.(5), and after having performed the integral over the monolayer, we find
\begin{equation}
\left\{
\begin{array}{ll}
\displaystyle 
\displaystyle E_{cS}(\kappa)&\displaystyle=-2\pi\frac{NQ^2}{A}\frac{1}{\kappa}\sum_{i=1}^N\exp(-\kappa\mid z_i\mid)\\
&\\
\displaystyle E_{SS}(\kappa)&\displaystyle=\pi\frac{N^2Q^2}{A}\frac{1}{\kappa}
\end{array}
\right .
\end{equation}
Therefore, in the small screening limit, we have 
\begin{equation}
E(\mbox{Yukawa};\kappa\rightarrow 0)=E_{cc}(\mbox{Coulomb})+2\pi\frac{NQ^2}{A}\sum_{i=1}^N\mid z_i\mid+o(\kappa)
\end{equation}
with $E_{cc}(\mbox{Coulomb})$ given by Eq.(15). Taking into account that the energy of Coulomb systems computed with Ewald sums or with Lekner sums agree very well \cite{f7}, then the result obtained in Eq.(23) is fully equivalent to the result obtained with Ewald sums. Thus, from an analytical point of view, Ewald and Lekner sums for quasi-two dimensional systems are fully equivalent. 

\section{Numerical tests for Lekner sums.}

As outlined before for Coulomb potentials \cite{f5,f6,f7}, the uses of Lekner sums to obtain interaction energies (or forces in molecular dynamics implementations) have to be done cautiously. For some configurations of particles, the convergence of sums may be very slow \cite{f7} and many contributions have to be included to obtain rather accurate results and to avoid bias in the sampling of the phase space \cite{f5}. For non-screened Coulomb interaction, the slow convergence rate stems from the first contribution in the right handed side of Eq.(15) ; the origin of this slow convergence rate is in the behaviour of the modified Bessel function K$_0$, or K$_1$ in molecular dynamics computations, as their argument tends to zero. In Lekner sums for Yukawa potentials, the same kind of difficulties is encountered since a similar contribution is included in the energy (the second contribution in Eq.(8)). There are mainly two methods to improve the convergence rate of these summations : the Lekner-cyclic \cite{f4,f7} and Lekner-Sperb \cite{f8,f9,f10} methods. The method derived by Sperb consists in deriving an alternative formula that converges faster than the original expression involving the Bessel functions, while the Lekner-cyclic method consists in using the symmetry of Eq.(16).\\
The Coulomb-like slow convergence in Lekner sums for Yukawa potentials is outlined below similarly to the study done in ref.\cite{f5}. We consider a pair of particles that carry charges $Q_1=+1$ and $Q_2=+1$, in a box with dimension $L_x=L_y=20\mbox{ }\sigma$ where $\sigma=1$ is a typical length scale of the system as a hard sphere diameter ; in these reduced units the energy at contact $E_{12}(r=\sigma)=1$. Particle 1 is located at (0,0,0) and particle 2 at $\bm{r}_{12}=(x_{12}, y_{12},z_{12})$. The difference  $\Delta E_{12}(\kappa ; p_c,m_c)$ is the difference between the particle-particle interaction energy computed by using $\Phi$ or $\Psi$ for the potential, it is defined by     
\begin{equation}
\displaystyle \Delta E_{12}(\kappa ; p_c,m_c) = Q_1Q_2\mbox{\large{(}} \Phi(\bm{r}_{12};L_x,L_y;p_c,m_c)-\Psi(\bm{r}_{12};L_x,L_y;p_c,m_c) \mbox{\large{)}}
\end{equation}
On Figure 1, we show a topographical representation of $\Delta E_{12}(\kappa ; p_c, 3)$ for several values of $\kappa$ and $p_c$, such as $-5\leq x_{12}\leq 5$ and $-5\leq y_{12}\leq 5$ and $z_{12}=0.1$, while on Figure 2, we show a surface representation of the same quantity for same configurations of the pair. The simulations box dimensions are $L_x=L_y=20\mbox{ }\sigma$, on Figures 1 and 2, we represent only situations where $\mid x\mid\leq 5$ and $\mid y\mid\leq 5$  to avoid irrelevant complication induced by the minimum image convention that should be used to compute the energy. The oscillatory behaviour of $\Delta E_{12}(\kappa ; p_c,m_c)$ has the same interpretation as for the Coulomb potential \cite{f5}.\\
Further numerical examinations of the analytical derivations done in the previous sectionshow that the slow convergence rate for these contributions to Lekner sums for Yukawa potentials may be handled by exactly the same means as it is done for Coulomb potentials. Thus, to implement the Lekner-cyclic method for systems with Yukawa interaction potentials one may proceed exactly as it is done for Coulomb potentials. Similarly to Eq.(28) in ref.\cite{f7}, the criterion to choose which one formula between $\Psi$ and $\Phi$ (cf. Eq.(16)) has to be used to compute the energy of the pair, may be taken as
\begin{center}
$\displaystyle \phi_{ij}= \left\{\begin{array}{ll}
        \Psi_{ij}(p_{c},m_{c})\t2\t2&\displaystyle\mbox{for}\t2\t2 \frac{|x_{ij}|}{L_x} > \frac{| y_{ij}|}{L_y}\\
        &\\
        \Phi_{ij}(p_{c},m_{c})\t2\t2&\displaystyle\mbox{for}\t2\t2 \frac{|x_{ij}|}{L_x} < \frac{| y_{ij}|}{L_y}
         \end{array}
        \right.$
\end{center}
and, similarly to Eq.(31) of ref.\cite{f7}, the truncation index $p_c$ for the sums over the cosine functions can be chosen such as  
\begin{center}
$ \displaystyle \sqrt{\mbox{\large{[}}\kappa^2+\frac{4\pi^2}{L_x^2}p_{c}^2(y_{ij},z_{ij})\mbox{\large{]}}\mbox{\large{[}}\mbox{\large{(}}{y_{ij}}+m L_y\mbox{\large{)}}^{2}+z_{ij}^{2}\mbox{\large{]}}}>19.$
\end{center}
As shown on Figures 1 and 2, the value of $p_c$ may be reduced significantly if $\kappa$ is large ; however, as it is shown in Table 2 and Eq.(47) of ref.\cite{f1}, for these large values of $\kappa$ a direct truncation of the potential and the minimum image convention are sufficient to obtain a good accuracy. Therefore, to implement the Lekner-cyclic method for Yukawa potentials, one may handle the Coulomb-like slow convergence behaviour of the sums exactly as it is done for non-screened Coulomb potentials.\\   
For Yukawa potentials, the Lekner sums derived in the previous section have an additional slow convergence behaviour : in Eq.(8), the first contribution in the right handed side involves a single summation over Bessel functions ;  as $\kappa\rightarrow 0$ and $\kappa\neq 0$, because of the asymptotic behaviour of K$_0(x)$ as $x\rightarrow 0$, a lot of contributions in the summation have to be included. As we will see below, this additional slow convergence behaviour is very penalising for implementations of Lekner sums for Yukawa potentials.\\
For a system of two Yukawa particles carrying charge $Q_1=+1$ and $Q_2=-1$, the Coulomb interaction energy can be obtained from Eq.(14) as $\kappa\rightarrow 0$. Since for this system, we have $Q_1+Q_2=0$, no background is needed to restore the electroneutrality of the system. Such test configurations have already been considered for Coulomb potentials in refs.\cite{f1,f3,f4,f5,f12,f13} that provide some reference points for the Coulomb limit, for these configurations we take $L_x=L_y=1$ as in original works.  On Figure 3, we give $E_{cc}(\mbox{Yukawa}; \kappa)$ as functions of $\kappa$ computed with Eqs.(8) and (9) inserted into Eq.(1), for configurations already considered in numerical tests of Lekner sums for Coulomb potentials and in numerical tests of Ewald sums for Yukawa potential \cite{f1}. On Fig.3, a thick horizontal line, for each reference configurations, indicates the limit obtained with a straightforward use of Lekner sums with $p_c=50$ and $m_c=4$. In Table 1, we report, for each of five configurations already considered for the Coulomb potentials in refs.\cite{f3,f4,f12,f13,f5}, the Coulomb limit ($\kappa=0$) and the values obtained for Yukawa potentials with : $\kappa=10^{-6}$, $p_c=50$, $m_c= 3$ and $50$ [Note : In reduced units and with $L_x=L_y=20$, a value as small as  $\kappa=10^{-6}$ for a box of side $L=1$ corresponds to a reduced screening parameter $\kappa^*=\kappa a = 5\times10^{-8}$ , cf. ref.\cite{f1}]. These data show that even by choosing a value as large as $m_c=50$ the Coulomb limit is not well reproduced. Values as large as $m_c=50$ for the truncation of the summations over the Bessel functions are not a reasonable choice, moreover such choice do not allow to reproduce with a sufficient accuracy the Coulomb limit.\\
This lost of accuracy of the Lekner sums for small $\kappa$, stems from the first contribution in the right hand side of Eq.(8) ; it is different from the slow convergence observed in Lekner sums for Coulomb potentials and increasing $p_c$, as it should be done for Coulomb potential, does not allow to achieve a better accuracy. Figure 4 shows that the increase of $p_c$ do not permit to obtain the Coulomb limit ($\kappa=0.0001$) with a good accuracy ;  for the configuration $(x_{12},y_{12},z_{12})=(0.4,0.4,0.1)$, considered in Fig.4, any value of $p_c>10$ give the same results for a $m_c$ fixed, but a strong dependence on $m_c$, at $p_c$ fixed, is observed. The same is true for the other test configurations. Examination of Eq.(8) allow to understand easily this independence on $p_c$ : the first contribution is independent of $p_c$.\\
To overcome this lost of accuracy for small $\kappa$ in Lekner sums by using Eqs.(8) and (9), we replace the first summation in Eq.(8) by its Taylor's expansion  
\begin{equation}
\begin{array}{ll}
\tilde{\Phi}(\bm{r}) &\displaystyle =\frac{4}{L_x}\sum_{p=1}^{+\infty}\cos\mbox{\large{(}}2\pi p \frac{x}{L_x}\mbox{\large{)}}\sum_{m=-\infty}^{+\infty}\mbox{ K}_0\mbox{\Large{[}}\sqrt{\mbox{\large{(}}\kappa^2+\frac{4\pi^2p^2}{L_x^2}\mbox{\large{)}}\mbox{\large{(}}(y+mL_y)^2+z^2\mbox{\large{)}}}\mbox{\Large{]}}\\
&\\
&\displaystyle +\frac{2\pi}{\kappa A}-\frac{\ln 2}{L_x}-\frac{1}{L_x}\ln\mbox{\huge{[}}\cosh\mbox{\large{(}}\frac{2\pi z}{L_y}\mbox{\large{)}}-\cos\mbox{\large{(}}\frac{2\pi y}{L_y}\mbox{\large{)}}\mbox{\huge{]}}+\kappa \frac{\pi z^2}{A}+o(\kappa^2)
\end{array}
\end{equation}
and self contributions in Eq.(9) have to be evaluated as
\begin{equation}
\begin{array}{ll}
\tilde{\Phi}_0 &\displaystyle =\frac{8}{L_x}\sum_{p=1}^{+\infty}\sum_{m=1}^{+\infty}\mbox{ K}_0\mbox{\huge{[}}m L_y\sqrt{\kappa^2+\frac{4\pi^2p^2}{L_x^2}}\mbox{\huge{]}}+\frac{2\pi}{\kappa A}+\frac{2}{L_x}\mbox{\Large{[}}\gamma+\ln\mbox{\Large{(}}\frac{L_y}{4\pi L_x}\mbox{\Large{)}}\mbox{\Large{]}}+\kappa+ o(\kappa^2)
\end{array}
\end{equation}
Thus, the particle-particle interaction energy is corrected as
\begin{equation}
\tilde{E}_{cc}(\mbox{Yukawa}; \kappa)=\frac{1}{2}\sum_{i=1}^{N}\sum_{i\neq j}^N Q_i Q_j \tilde{\Phi}(\bm{r}_{ij})+\frac{1}{2}\sum_{i=1}^{N}Q_i^2\tilde{\Phi}_0
\end{equation}
More precisely, the particle-particle interaction energy, computed with the modified Lekner sums of Eqs.(25) and (26), is
\begin{equation}
\begin{array}{ll}
\displaystyle \tilde{E}_{cc}(\mbox{Yukawa}; \kappa) &\displaystyle = \frac{2}{L_x}\sum_{i,j}\mbox{}'Q_i Q_j\sum_{p=1}^{p_c}\sum_{m=-m_c}^{m_c}\cos\mbox{\large{(}}2\pi p \frac{x_{ij}}{L_x}\mbox{\large{)}}\mbox{ K}_0\mbox{\Large{[}}\sqrt{\mbox{\large{(}}\kappa^2+\frac{4\pi^2p^2}{L_x^2}\mbox{\large{)}}\mbox{\large{(}}(y_{ij}+mL_y)^2+z_{ij}^2\mbox{\large{)}}}\mbox{\Large{]}}\\
&\\
&\displaystyle -\frac{1}{2L_x}\sum_{i\neq j}Q_i Q_j \ln\mbox{\huge{[}}\cosh\mbox{\large{(}}\frac{2\pi z_{ij}}{L_y}\mbox{\large{)}}-\cos\mbox{\large{(}}\frac{2\pi y_{ij}}{L_y}\mbox{\large{)}}\mbox{\huge{]}}-\frac{1}{2L_x}\mbox{\large{(}}\sum_{i=1}^{N}Q_i\mbox{\large{)}}^2\ln 2\\
&\\
&\displaystyle +\frac{1}{L_x}\mbox{\large{(}}\sum_{i=1}^{N}Q_i^2\mbox{\large{)}}\mbox{\large{[}}\gamma + \ln\mbox{\large{(}}\frac{L_y}{L_x}\mbox{\large{)}}-\frac{1}{2}\ln(8\pi^2) \mbox{\large{]}} +\frac{\pi}{A}\mbox{\large{(}}\sum_i Q_i\mbox{\large{)}}^2\frac{1}{\kappa}\\
&\\
&\displaystyle+\frac{\pi}{2}\frac{\kappa}{A}\sum_{i\neq j}Q_i Q_j z_{ij}^2+\frac{\kappa}{2}\sum_{i}Q_i^2+o(\kappa^2)
\end{array}
\end{equation}
This partial expansion of the particle-particle interaction energy allows to obtain the correct Coulomb limit as $\kappa$ is small, while keeping $m_c$ to a reasonable value. This is illustrated on Table 1 and on Figure 5, where $\tilde{E}_{cc}(\mbox{Yukawa}; \kappa)$, computed with $m_c=3$ and the modified Lekner sums, is plotted for the configuration $(0.4, 0.4,0.1)$. As shown on the inset of Figure 5, the corrected value for the energy may be obtained with accuracy for $\kappa < 0.1$.\\
For values of the screening parameter $\kappa$ between 0.1 and 2, that are interesting values of the screening parameter for many systems, the accuracy of the corrected value $\tilde{E}_{cc}(\mbox{Yukawa}; \kappa)$ has to be improved by adding high order contributions to $\tilde{\Phi}(\bm{r})$ and $\tilde{\Phi}_0$ to keep a reasonable value for $m_c$. Contributions in $\kappa^2$ and $\kappa^3$ are given by
\begin{equation}
\left\{
\begin{array}{ll}
\displaystyle C_2(\bm{r})\kappa^2&\displaystyle =\frac{2L_y^2}{L_x} \mbox{\huge{[} }\frac{1}{8\pi^2}\mbox{ Re} \mbox{\Large{[}}\mbox{Li}_3(\Omega)\mbox{\Large{]}}-\frac{z}{\pi L_y}\mbox{ Re} \mbox{\Large{[}}\mbox{Li}_2(\Omega) \mbox{\Large{]}}-\frac{\pi}{6}\frac{z^3}{L_y^3} \mbox{ \huge{]}}\kappa^2\\
&\\
\displaystyle C_3(\bm{r})\kappa^3&\displaystyle =\frac{\pi}{12}\frac{z^4}{A}\kappa^3
\end{array}
\right .
\end{equation}
with $\mbox{Re}(z)$ the real part of the complex number $z$, the polylogarithm functions  $\mbox{Li}_{\nu}(z)$ and $\Omega$ defined by
\begin{center}
$\Omega=\exp(-2\pi(z+iy)/L_y)\mbox{      }$ and $\mbox{      }\displaystyle \mbox{Li}_{\nu}(\Omega)=\sum_{p=1}^{\infty}\frac{1}{p^{\nu}}\Omega^p$
\end{center}
One has also to add to $\tilde{\Phi}_0$ the contributions
\begin{equation}
\left\{
\begin{array}{ll}
\displaystyle C_2^0\kappa^2&\displaystyle = -\frac{1}{12} L_x \mbox{\large{[}}1+3\frac{\zeta(3)}{\pi^2}\frac{L_y^2}{L_x^2}\mbox{\large{]}}\kappa^2\\
&\\
\displaystyle C_3^0\kappa^3&\displaystyle =\frac{1}{3} L_x^2\kappa^3
\end{array}
\right .
\end{equation}
where $\zeta(z)$ is the Riemann's Zeta function defined by
\begin{center}
$\displaystyle \zeta(z)=\sum_{p=1}^{\infty}\frac{1}{p^z}$
\end{center}
For $\kappa >2$, the particle-particle interaction energies may be evaluated with a good accuracy by using Eqs.(8) and (9) with $m_c=3$. From data on Figure 5 and by using the original Lekner sums in Eq.(8) and (9) at $\kappa=2$ and with $p_c=50$, we have for $m_c=3$, $E_{cc}=-0.880722$ and $m_c=50$, $E_{cc}=-0.880976$.\\
It is worthwhile to note that an alternative summation for the first contribution in the right hand side of Eq.(8) has been derived in the appendix of ref.\cite{f15} (cf. Eq.(A4)). In notations of the present paper, it becomes
\begin{equation}
\begin{array}{ll}
\displaystyle\sum_{m=-\infty}^{+\infty}\mbox{ K}_0\mbox{\Large{[}}\kappa\sqrt{(y+mL_y)^2+z^2}\mbox{\Large{]}}&\displaystyle=2\pi\sum_{m=1}^{+\infty}\frac{\cos(2\pi m y/L_y)}{\sqrt{(2\pi m)^2+L_y^2\kappa^2}}\exp\mbox{\large{(}}-z\sqrt{(2\pi m/L_y)^2+\kappa^2} \mbox{\large{)}}\\
&\\
&\displaystyle +\frac{\pi}{\kappa L_y}e^{-\kappa z}
\end{array}
\end{equation}
From this equation, we recover easily Eq.(10) as $\kappa\rightarrow 0$ ; if $\kappa\neq 0$, the summation over $m$ in the right hand side of Eq.(31) is slowly convergent and a lot of contributions have to be included to obtain a good accuracy, especially if $z$ is small ; in agreement with Figures 3 and 4.

\section{Conclusion.}

In view of the numerical tests performed in the previous section and since Lekner sums are not more efficient than Ewald sums : for Yukawa interaction potential in quasi-two dimensional systems, it is highly recommended to use the Ewald sums derived in ref.\cite{f1}.

\newpage
\vspace{.5cm}

\newpage
\listoftables
\normalsize{\bf Table 1 :} Numerical tests of the Coulomb limit of the Lekner sums for the Yukawa potentials. The configuration of the pair of particles is defined by $(x_{12},y_{12},z_{12})$ particle 1 with a charge $Q_1=+1$ located at (0,0,0) and particle 2 with a charge $Q_2=-1$ located at $(x_{12},y_{12},z_{12})$. The values of particle-particle Coulomb energies are extracted from original works by others. These energies have been obtained with the Lekner sums for Coulomb potential with truncations indexes $(p_c,m_c)$ indicated in the Table and they agree with the values computed in ref.\cite{f5} with $p_c=30$ and $m_c=3$. Evaluations of particle-particle interaction energies in columns {\it Direct\/}, refer to evaluations performed by using straightforwardly the Lekner sums for Yukawa potentials given by Eqs.(1,8) and (9), and  with $p_c=50$ and $m_c$ as indicated in the Table. The term {\it Modified\/} refers to evaluations of the energies of pairs with Eq.(28) that uses Taylor's expansions, and with $p_c=50$ and $m_c$ as indicated in the Table (see also Figure 3 and 5). In the column {\it Ewald\/}, we give the values of the energy obtained for $\kappa=1.$ by using Ewald sums with $\alpha=6.$ and $k\times k = 16\times 16$ (see ref.\cite{f1} for the numerical tests of the Ewald sums for Yukawa potentials). 

\footnotesize
\newpage 
\begin{center}
\begin{tabular}{|c|cc|ccccc|c|}
\hline
\hline
&&&&&&&&\\
Configurations                     &                        & Coulomb            &               &  Direct            & Modified            & Direct               & Modified & Ewald \\ 
 and References & $(p_c,m_c)$ &   Limit       & $m_c$ &    $\kappa = 10^{-6}$ &   $\kappa = 10^{-6}$ &  $\kappa =1.$ & $\kappa =1.$ & $\kappa = 1.$\\
$(x_{12},y_{12},z_{12})$&&&&&&&&  \\
&&&&&&&& \\
\hline
\hline
&&&&&&&&\\
(0.5,0.5,0.) & analytical & -2.28472 & 3  & -2.14344  & -2.28472 & -1.41986 & -1.29568 & -1.4307\\
\cite{f4,f12} &  &                  & 50 & -2.27482 &                   & -1.43067 & &\\
&&&&&&&&\\
(0.1,0.1,0.1) & (14,3) & -5.77211 &   3 & -5.77212 & -5.77212 & -4.87007 & -4.77663 & -4.87041\\
\cite{f12,f13} &    &               & 50 & -5.77212 &  & -4.87041& & \\
 &&&&&&&&\\
(0.,0.,0.25)  & (10,3)  & -3.72483  & 3   & -3.76028 & -3.72483 & -2.98415 & -2.9022 &-2.98362\\
   \cite{f12}  &             &                    & 50 & -3.72731 &                  & -2.98362 & & \\
&&&&&&&&\\
(-0.25,-0.15,-0.2) & (10,3) & -2.82156  & 3   & -2.83153 & -2.82157 & -2.04423 & -1.94915& -2.04483\\
    \cite{f12}           &             &                   & 50 & -2.82226 &                  & -2.04483 & &\\
&&&&&&&&\\
(0.4,0.4,0.1) & not  & -2.28608 & 3   & -2.20082 & -2.28609 & -1.44881 & -1.32721& -1.45555\\
  \cite{f13}    & reported &          & 50 & -2.28015 &                  & -1.45555 & &\\
&&&&&&&&\\
\hline
\hline
\end{tabular}
\end{center}

\begin{center}
{\begin{quote}\item[\large\underline{\bf{Table 1}} M. Mazars, Yukawa II : Lekner sums.]\end{quote}} 
\end{center}

\normalsize

\newpage
{\Large\bf List of Figures}\\[0.2in]

\normalsize{\bf Figure 1:} Representation of $\Delta E_{12}(\kappa ; p_c,m_c)$ as contour plot for several values of $\kappa$ and $p_c$. For the six figures, the particle 1 is located at (0,0,0) and particle 2 at $(x,y,z)$ ; the common parameters for all figures are : $Q_1=Q_2=+1$, $L_x=L_y=20$, $-5\leq x\leq 5$,  $-5\leq y\leq 5$, $z=0.1$ and $m_c=3$. For each figure, 20 isopotentials are represented and the values of $\Delta E_{12}$ are contained between -0.5 and 0.5 (i.e. isopotentials are separated by 0.05 in reduced energy unit). For each figure, the value of $\kappa$ and $p_c$ are indicated on the figures. For $\mid x\mid >5$ or $\mid y\mid >5$, computations have not been done because of irrelevant complications induced by the minimum image convention.

\normalsize{\bf Figure 2:}  Surface representation of $\Delta E_{12}(\kappa ; p_c,m_c)$ with the same parameters as in Figure 1. The values of  $\Delta E_{12}$ are contained between -0.5 and 0.5.

\normalsize{\bf Figure 3:} Representation of $E_{cc}(\mbox{Yukawa}; \kappa)$ for four configurations of a pair of Yukawa particles as functions of $\kappa$. $E_{cc}(\mbox{Yukawa}; \kappa)$ are computed by using equations (8) and (9) inserted in (1), with $p_c=50$ and $m_c=4$. The particle 1 carries a charge $Q_1=+1$ and is located at (0,0,0) ; the particle 2 carries a charge $Q_2=-1$ and four positions of the particle 2 are considered. These configurations have already been considered in some previous works \cite{f5,f12,f13} (see also Table 1). For each configuration, the value of the Coulomb limit obtained from Eq.(1) with $p_c=50$ and $m_c=4$ is indicated by a thick red horizontal line and limiting values are explicitly given. The differences with the values obtained by using Lekner sums for non-screened Coulomb interactions are given in Table 1.  

\normalsize{\bf Figure 4:}  Representation of $E_{cc}(\mbox{Yukawa}; \kappa)$ computed by using equations (8) and (9) inserted in (1) for a pair of Yukawa particles as function of $m_c$ for three values of $p_c$. The configuration of the pair is defined by : Particle 1, $Q_1=+1$ located at (0,0,0) ; Particle 2,  $Q_2=-1$ located at (0.4,0.4,0.1) \cite{f5,f13}. The value of the screening parameter is $\kappa = 0.0001$, close to the Coulomb limit.

\normalsize{\bf Figure 5:} Representation of $E_{cc}(\mbox{Yukawa}; \kappa)$ as function of $\kappa$ for the configuration $(0.4,0.4,0.1)$ of particle 2. Curves with several values of $m_c$ in Eq.(1) and for a value $m_c=3$ in Eq.(28), that allows to  overcome the lost of accuracy of Eq.(1) for small $\kappa$, are represented ; for each curve $p_c=50$. The inset shows the values obtained for $0.1<\kappa<1$ where Taylor's expansions given in Eqs.(25) and (26) become inaccurate.

\newpage
\begin{figure}[htbp]
\begin{center}
\centerline{\includegraphics[width=6.in]{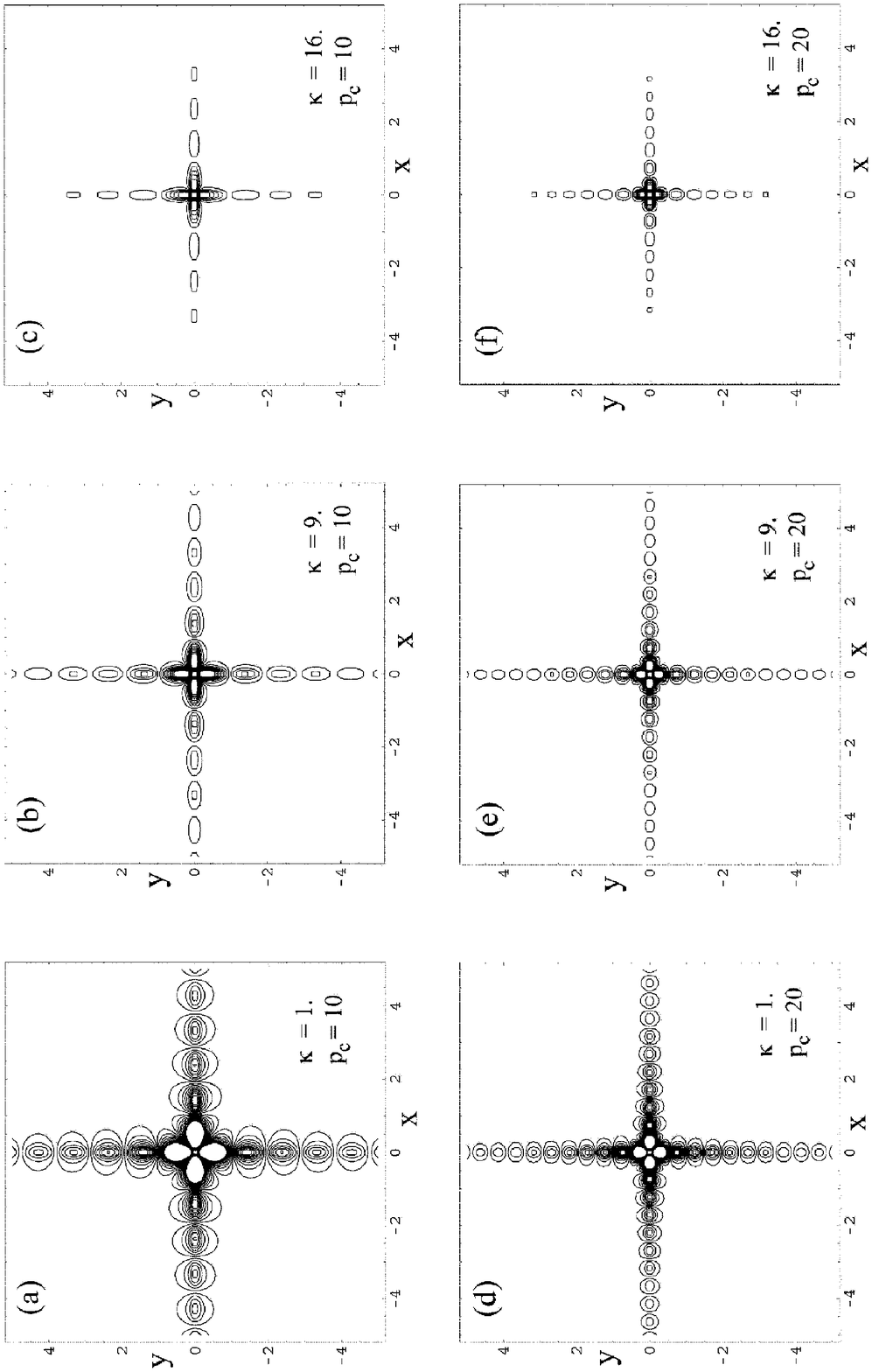}}
\caption{{\bf }  \large M. Mazars, Yukawa II : Lekner sums.}
\end{center}
\end{figure}

\newpage
\begin{figure}[htbp]
\begin{center}
\centerline{\includegraphics[width=4.3in]{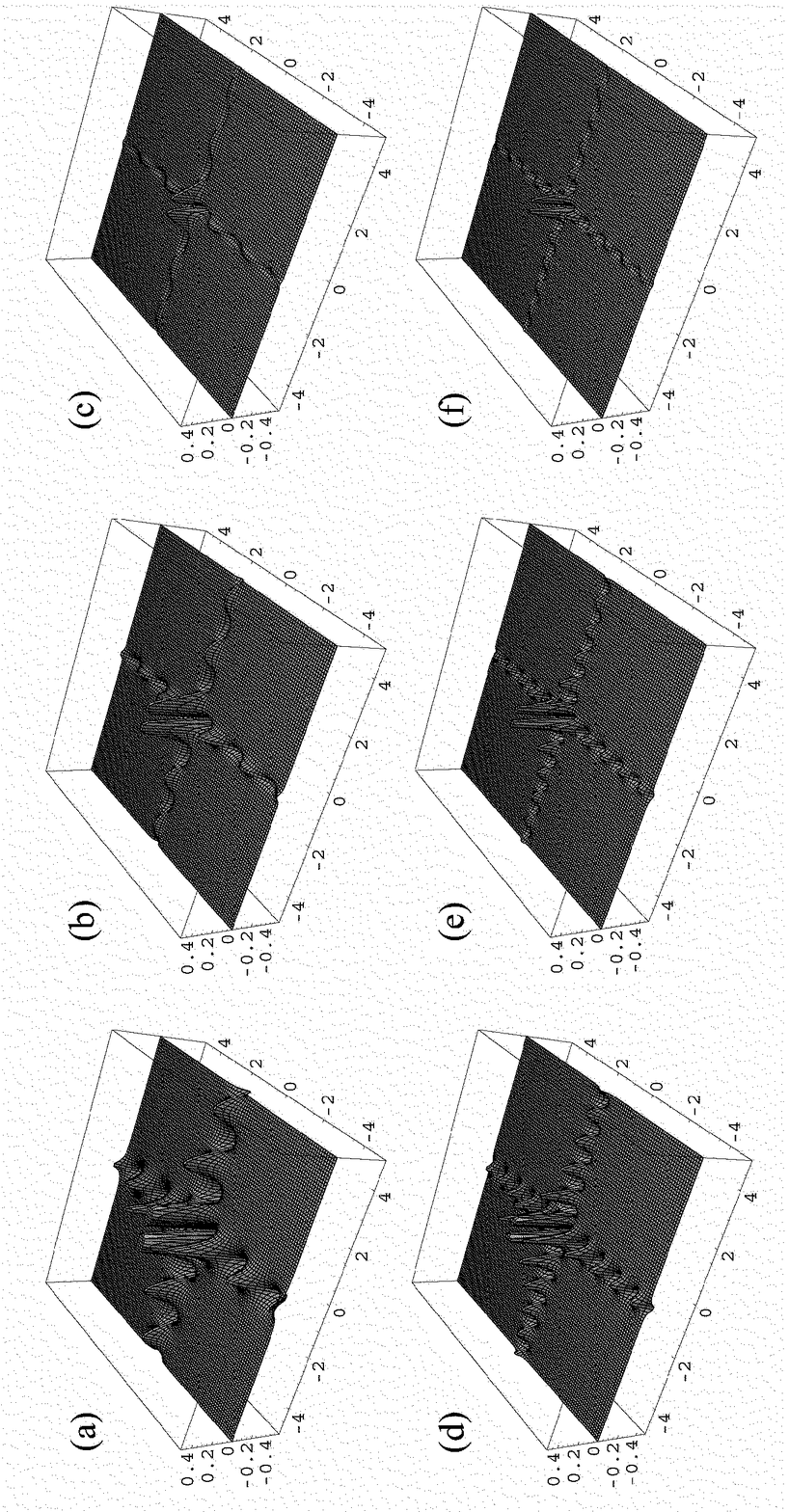}}
\caption{{\bf }  \large M. Mazars, Yukawa II : Lekner sums.}
\end{center}
\end{figure}

\newpage
\begin{figure}[htbp]
\begin{center}
\centerline{\includegraphics[width=7.3in]{MazarsTMPH-2007-0091Fig3.eps}}
\caption{{\bf }  \large M. Mazars, Yukawa II : Lekner sums.}
\end{center}
\end{figure}

\newpage
\begin{figure}[htbp]
\begin{center}
\centerline{\includegraphics[width=7.3in]{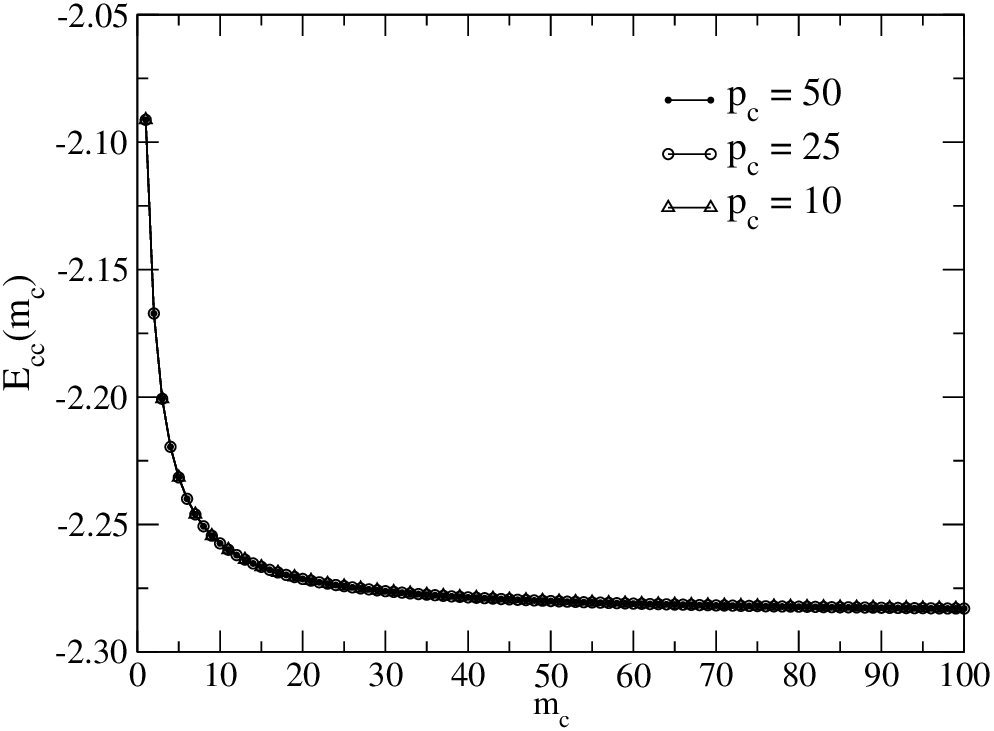}}
\caption{{\bf }  \large M. Mazars, Yukawa II : Lekner sums.}
\end{center}
\end{figure}

\newpage
\begin{figure}[htbp]
\begin{center}
\centerline{\includegraphics[width=7.3in]{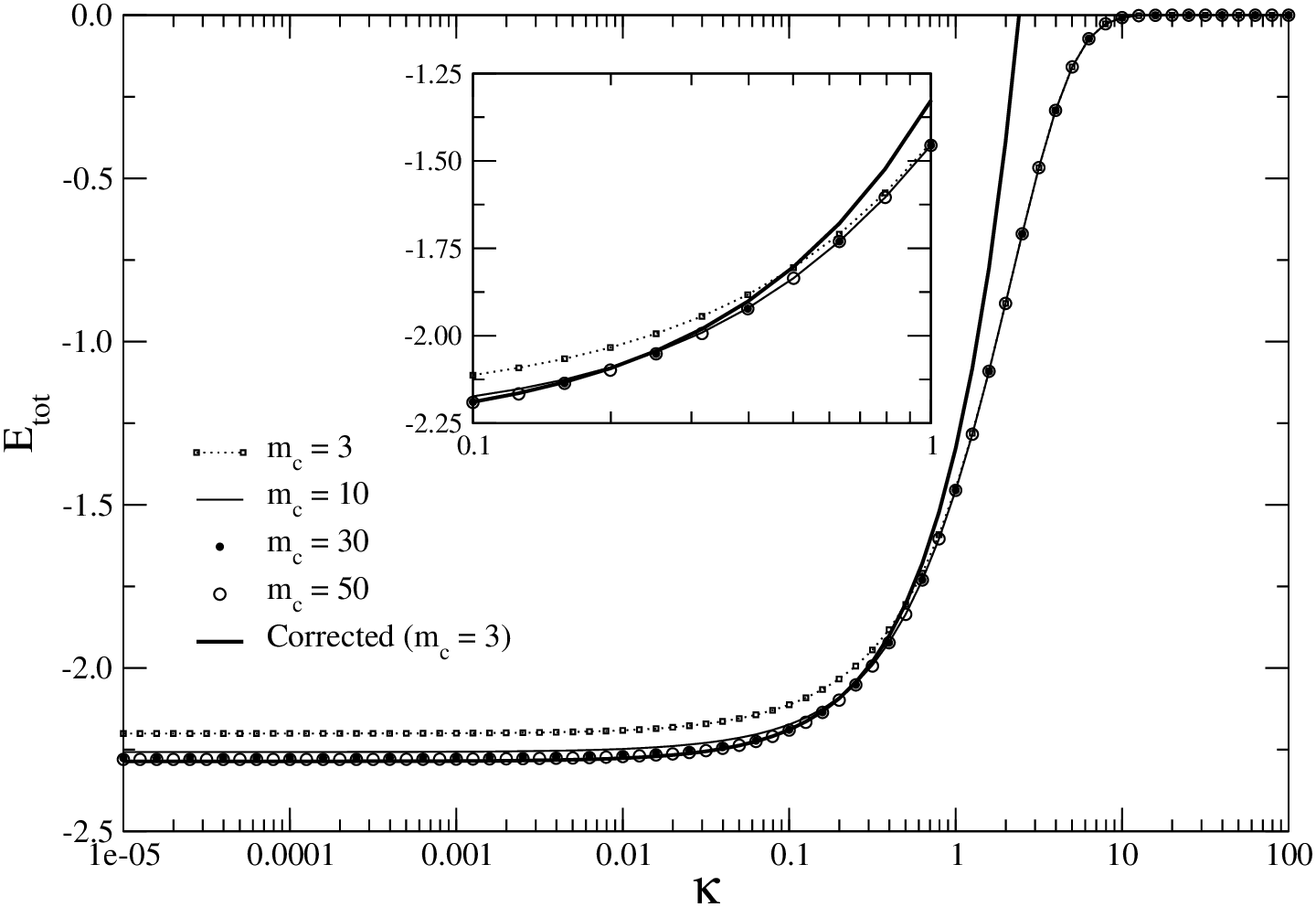}}
\caption{{\bf }  \large M. Mazars, Yukawa II : Lekner sums.}
\end{center}
\end{figure}

\end{document}